# Development of Smartphone Application for Evaluation of Passenger Comfort

Juraj Machaj[1][0000-0002-7544-8796], Peter Brida[1][0000-0002-5442-9246], Ondrej Krejcar[2][0000-0002-5992-2574], Milica Petkovic[3][0000-0002-1861-2668] and Quingjiang Shi[4]

[1] University of Zilina, Faculty of Electrical Engineering and Information Technology, Department of Multimedia and Information-Communication Technology, Univerzitna 1, 010 26 Zilina, Slovakia

[2] Center for Basic and Applied Research, Faculty of Informatics and Management, University of Hradec Kralove, Rokitanskeho 62, Hradec Kralove, 500 03, Czech Republic

[3] Faculty of Technical Sciences, University of Novi Sad, Trg Dositeja Obradovića 6, 21000 Novi Sad, Serbia

[4] School of Software Engineering, Tongji University, 4800 Cao an Road, 201804, Shanghai, China

{juraj.machaj; peter.brida}@feit.uniza.sk;
ondrej.krejcar@uhk.cz; milica.petkovic@uns.ac.rs;
shiqj@tongji.edu.cn

**Abstract.**
Nowadays, smartphones are not utilized for communications only. Smartphones are equipped with a lot of sensors that can be utilized for different purposes. For example, inertial sensors have been used extensively in recent years for measuring and monitoring performance in many different applications. Basically, data from the sensors are utilized for estimation of smartphone orientation. There is a lot of applications which can utilize these data. This paper deals with an algorithm developed for inertial sensors data utilization for vehicle passenger comfort assessment.

## 1    Introduction

In recent time transport systems are facing significant changes in the behaviour of passengers. This might be caused by increased awareness about pollution generated by cars, as well as the development of new services allowing shared mobility. The idea of shared mobility is that a group of passengers share a single vehicle for their journey.

With the development of intelligent transport system, there is a large amount of data that can be used to improve provided services and thus comfort levels of passengers. However, passengers comfort levels can be decreased by aggressive driving.

In order to keep the passengers safe and satisfied drivers of shared cars, or any other service like UBER, DiDi etc., should not drive aggressively and dangerously. If such drivers are present in the system, they should be identified and denied to provide further service. This can be done by collecting feedback from the passengers. However, passengers might be in a hurry or not willing to provide feedback in the mobile application.

In this paper, we propose a solution for the automatic evaluation of passenger comfort, which can run on a smartphone device. The motivation is to allow the collection of information about driver behaviour on a widely used platform. A similar system was used to analyze the impact of different materials used to build transport infrastructure on the comfort level of passengers in [1, 2]. However, this system was based on a device specifically designed to provide the required data. Since this specialized device has to be implemented in all tested vehicles, its implementation on a greater scale would be financially ineffective. Therefore, we proposed a solution to use data from the Inertial Measurement Unit (IMU) inbuilt in the majority of current smartphones. This should allow wide adoption of the solution without any hardware costs since smartphones are used ubiquitously.

The rest of the paper is organized as follows, the next section will describe IMU and individual sensors that could be used for the purpose of comfort level assessment, Section 3 will introduce the proposed solution. Analysis of the achieved results will be provided in Section 4 and Section 5 will conclude the paper.

## 2    Inertial measurement units

The inertial measurement unit consists of multiple sensors that are used to detect and measure external forces applied to the device. Typically, IMU consists of Accelerometer, Gyroscope and Magnetometer sensors [3 - 12]. These sensors are used to measure acceleration using gravitational force, orientation and variations of a magnetic field, respectively. Variations of a magnetic field can be used to estimate heading of the device with respect to North, however, measurements are affected by disturbances of a magnetic field caused by metal objects. Since in the proposed solution data from accelerometer and gyroscopes are used, these will be described in more detail.

### 2.1    Accelerometer

The accelerometer can be used to measure acceleration using inertial parameters of objects. In case when an object is still, the accelerometer will report acceleration equal approximately to 9.81 $ms^{-1}$ which represent the gravitational force of the Earth. The accelerometer cannot separate gravitational force from other acceleration forces affecting the device, therefore the gravitational force has to be compensated in the application.

Accelerometers implemented in the smartphones are using micro-electro-mechanical systems (MEMS). Their accelerometers can measure acceleration only in one direction, perpendicular to axis od the matrix. Therefore, in order to provide measurements in all axes, three MEMS accelerometers must be implemented.

## 2.2 Gyroscope

Gyroscope is a device able to sense an angular velocity of the object. A triple-axis gyroscope can measure rotation around three axes: x, y, and z. There are MEMS gyros implemented in smartphones because the quality is sufficient at a reasonable price. The gyroscope sensor within the MEMS is very small (between 1 to 100 micrometres). If the gyro is rotated, a small resonating mass is shifted as the angular velocity changes. The principle is based on Coriolis force

$$\mathbf{F_{cor}} = 2m(\boldsymbol{\omega} \times \mathbf{v}), \tag{1}$$

where $m$ is the weight of a vibrating element, $\boldsymbol{\omega}$ is angular velocity, $\mathbf{v}$ is the speed of the element and $\mathbf{F_{cor}}$ is Coriolis force.

## 2.3 Measurement errors

The measured data from the sensors are influenced by measurement errors. The physical properties of these sensors change over time which results in different characteristics over time. Following phenomenon should be considered when IMU measurements are used:

- Repeatability: It is the ability of the sensor to deliver the same output for the same repeated input, assuming all other conditions are the same
- Stability: It is the ability of the sensor to deliver the same output, over time, for the same constant input.
- Drift: It is the change of the output over time (zero drift is the change over time with no input).
- Bias: For a given physical input, the sensor outputs a measurement, offset by the bias.
- Noise: the random fluctuation of the output data, can be reduced by filters.

For example, when input rotation is null, the output of the gyro could be nonzero. The equivalent input rotation detected is the bias error. Depending on sensor usage the internal sensor biases may increase over time. In order to compensate these errors, the gyroscope must be calibrated. This is usually done by keeping the gyroscope still and zeroing all the readings.

# 3 Proposed system

In the proposed system data from both gyroscope and accelerometer sensors are used in order to detect and events that have a negative impact on passenger comfort level, i.e. detection of high acceleration, hard braking, uneven roads, etc. Data from the sensors are, however, affected by noises coming from multiple sources. For example, in Fig. 1 raw data from the accelerometer in the vertical axis during the ride on a road are presented. Under ideal conditions, these data should have constant value as it impacts of gravitational force. However, the acceleration value fluctuates quite significantly.

This might be caused by multiple factors, one of them being an uneven road which might have an impact on comfort levels of passengers. Other parameters that can cause fluctuations of the acceleration inv vertical axis include vibrations from the engine, the setting of the car suspension that should reduce the impact of the uneven road but also introduce some secondary fluctuations.

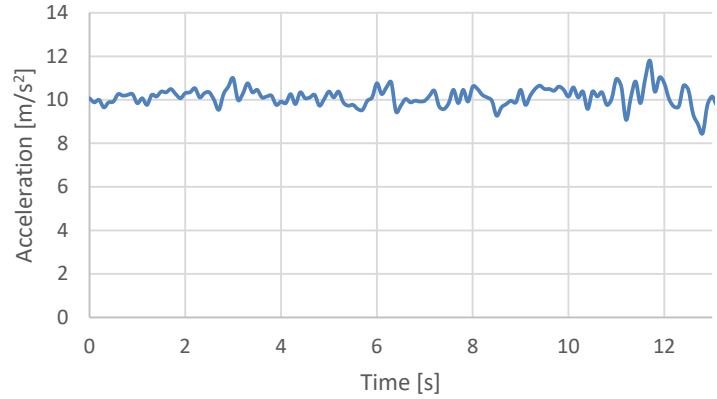

**Fig. 1.** Vertical acceleration from the raw data

Therefore, to reduce detection error low pass filter was used to smoothen raw data from the IMU. The implemented filter can be described as

$$out(i) = out(i-1) + \alpha\big(input(i) - out(i-1)\big), \tag{2}$$

where α is defined as the time constant of filter t divided by the sum of the time constant and time of individual accelerations provided by sensor dT:

$$\alpha = \frac{t}{t+dT}. \tag{3}$$

Since low past filter should remove high-frequency noises the filtered value should include only changes in acceleration caused by uneven road, the acceleration on vertical axes after filtration is shown in Fig. 2. However, the data are still affected by noises generated by the sensor itself as described in the previous chapter.

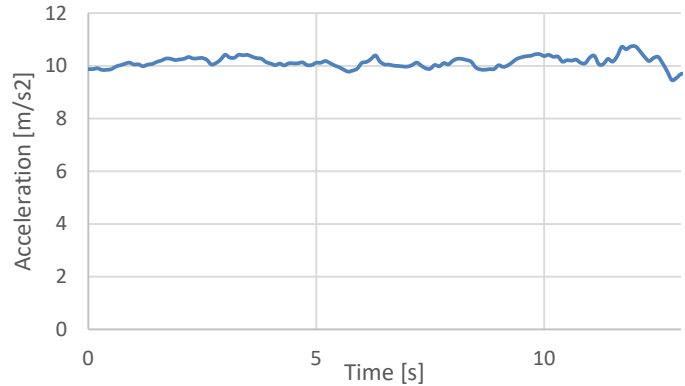

**Fig. 2.** Vertical acceleration from the filtered data

On the other hand, data from the horizontal plane are not affected by gravitational force in case that vehicle is not going uphill or downhill. In horizontal plane data from 2 axes are collected. Accelerometer data from x-axis represent the change in the speed of the vehicle, as it measures positive or negative acceleration in case of accelerating or slowing down. The data from the y-axis, on the other hand, represent a centrifugal force that is affecting vehicle while driving thru a curve.

These data are not affected by the quality of the road, however, can provide crucial information about the behaviour of the driver. On the other hand, the data are affected by vibrations caused by the vehicle engine. Therefore, the same filtering approach as for vertical axis was applied on both axes in the horizontal plane. In case that driver is driving more aggressively, forces affecting accelerometer will be higher and thus can provide feedback about reckless driving.

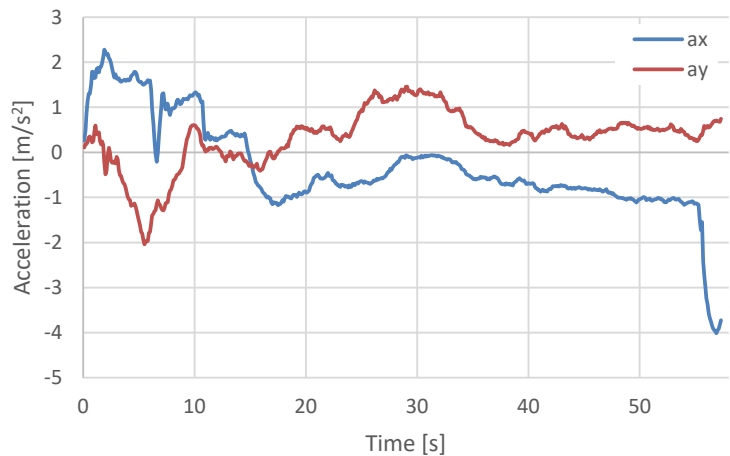

**Fig. 3.** Acceleration from both axes in the horizontal plane

In Fig. 3 accelerometer data from both axes in the horizontal plane are shown. Data ax is from x-axis that shows acceleration or braking of the car, and ay represents the impact of centrifugal force when the car is driving thru the curve.

When it comes to the data from the gyroscope, the rotation on vertical z-axis gives information about the rotation of the vehicle when driving thru the curve. Rotation on y-axis give information about ascend or descend of the vehicle and rotation on x-axis gives information about uneven roads.

The axes should be fixed to the vehicle and not to the smartphone device, that might have axes shifted due to the position of the device within the vehicle. Therefore, it is required to map axes of the smartphone to axes of the vehicle. This can be done using a transformation matrix for conversion between different coordinate systems. The transformation matrix is defined as:

$$\begin{bmatrix} x^v \\ y^v \\ z^v \end{bmatrix} = \left( R_z^{-1} \left( R_y^{-1} \right) R_x^{-1} \begin{bmatrix} x \\ y \\ z \end{bmatrix} \right),$$  (4)

where [x, y, z] are values on axes of the smartphone, $x^v$, $y^v$, $z^v$ are axes of vehicle and $R_x$, $R_y$, $R_z$ are rotation matrices.

In the first step vertical alignment is performed. To perform vertical alignment 2 approaches were implemented. The first approach is angular velocity method. This method can be applied when the vehicle is going thru the curve. In such case the angular velocity is given by the velocity of vehicle and diameter of the curve and can be expressed as:

$$\omega = \frac{v}{r}.$$  (5)

When axes of the gyroscope are mapped to the axes of the car, the vertical z-axis should have significantly higher angular velocity than any other axis. However, if the gyroscope is not aligned with axes of the vehicle the fact that car is turning can be determined based on total angular velocity, given by:

$$\omega_t = \sqrt{\omega_x^2 + \omega_y^2 + \omega_z^2}.$$  (6)

Components of the rotation matrix R used for alignment in the vertical axis can be calculated using:

$$\begin{bmatrix} R_x^v \\ R_y^v \\ R_z^v \end{bmatrix} = \frac{1}{\sqrt{\omega_x^2 + \omega_y^2 + \omega_z^2}} \begin{bmatrix} \omega_x \\ \omega_y \\ \omega_z \end{bmatrix}.$$  (7)

In Fig. 4 gyroscope data transformed into vehicle coordination system using equation (7) are shown.

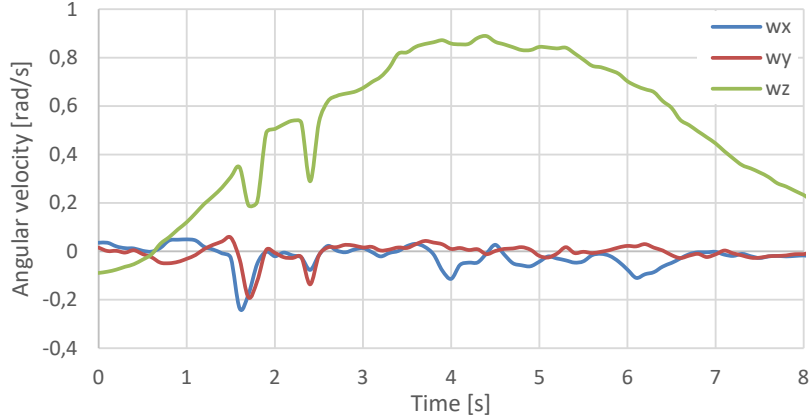

**Fig. 4.** Angular velocity after alignment with the vehicle coordination system

From the figure, it is obvious that angular velocity on the z-axis is significantly higher than angular velocities recorded on horizontal axes, which are oscillating around 0.

Vertical alignment using angular velocity is suitable when the vehicle is moving on the curve, however, in a situation when the vehicle is not moving or is moving on the straight section of road it is required to use a different approach for alignment of the vertical axis. Under these conditions, gravitational alignment method can be used. In this method data from accelerometers are used. Moreover, it is assumed that the only force that is affecting the IMU measurements is a gravitational force, therefore, this approach work when the vehicle is still or is moving with constant speed on a straight section of the road. In such a case, the total acceleration is given by:

$$a_t = \sqrt{a_x^2 + a_y^2 + a_z^2},$$ (8)

and should be equal to the impact of gravitational force. However, even in a case when the vehicle is moving on the constant speed the acceleration fluctuates due to the uneven surface of the road. Unfortunately, the frequency of this fluctuation cannot be determined, therefore it is not possible to remove it using low pass filter. Anyhow, the rotation matrix for transformation between coordinate systems using gravitational method can be defined as:

$$\begin{bmatrix} R_x^v \\ R_y^v \\ R_z^v \end{bmatrix} = \frac{1}{\sqrt{A_x^2 + A_y^2 + A_z^2}} \begin{bmatrix} A_x \\ A_y \\ A_z \end{bmatrix},$$ (9)

where $A_x$, $A_y$, $A_z$ are values of acceleration measured by accelerometers on individual axes.

Horizontal alignment can be performed under two scenarios as well. In the first scenario, the vehicle is accelerating, and the acceleration vector is the same as the direction of the movement. On the other hand, if the vehicle is driving thru the curve on constant speed, then acceleration vector will be perpendicular to the direction of vehicle movement and therefore it will be mapped on the *y*-axis of the vehicle.

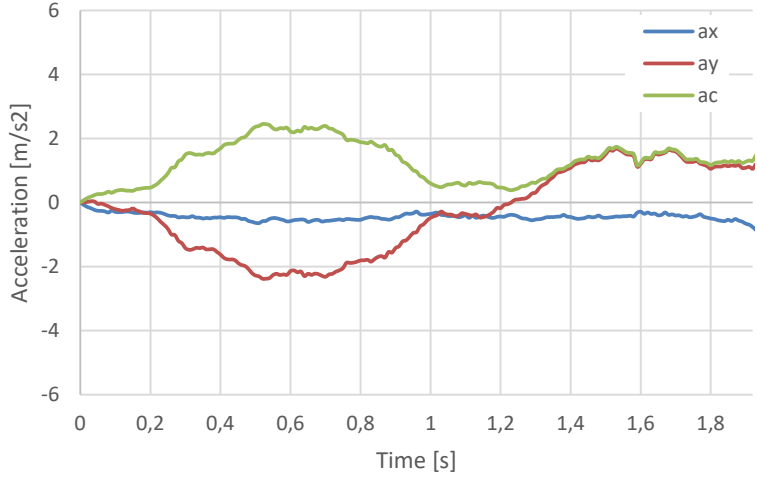

**Fig. 5.** Measured and calculated acceleration in the horizontal plane

In Fig. 5 real data measured by accelerometers in horizontal plane $a_x$ and $a_y$ are presented together with combined acceleration ac given by:

$$a_c = \sqrt{a_x^2 + a_y^2} \, .$$ (10)

To align horizontal axes, it is required to estimate forces caused by acceleration (or breaking) of the vehicle and map these to *x*-axis and acceleration caused by centrifugal force that should be mapped to the *y*-axis. From the figure, it can be seen that the vehicle is moving on the constant speed since acceleration on the *x*-axis is close to 0. Then, the value of $a_c$ is close to the absolute value of acceleration on the *y*-axis.

The other option is to align the horizontal axes during vehicle acceleration. The basic assumption is that the vehicle is moving straight forward. In such case value of ac is actually the value of acceleration in the x-axis.

After the data extraction, reduction of data dimensions is performed using Principal Component Analysis (PCA). The goal is to find interconnection between variables. In the first step, the covariance matrix is calculated. In our case relation between vertical and horizontal acceleration is investigated using:

$$\text{cov}(x, y) = \frac{\sum_{i=1}^{n}(x_i - \bar{x})(y_i - \bar{y})}{(n-1)} = \text{E}[xy] - \text{E}[x]\text{E}[y].$$ (11)

For multiple variables the covariance can be calculated for each pair individually and then covariance matrix can be formed as follows:

$$\mathbf{C} = \begin{pmatrix} cov(x,x) & cov(x,y) & cov(x,z) \\ cov(y,x) & cov(y,y) & cov(y,z) \\ cov(z,x) & cov(z,y) & cov(z,z) \end{pmatrix}. \tag{12}$$

Using equations above it is possible to calculate eigenvalues and eigenvectors. The outcome of PCA is eigenvector **m**, which represent direction, in which the covariance is the highest, i.e. direction of vehicle movement.

Therefore, it is possible to estimate angle $\phi$ using:

$$\phi = atan2(m_2, m_1), \tag{13}$$

where $m_2$ and $m_1$ are values of eigenvector **m**.

## 4 Testing and evaluation

In order to evaluate the comfort level of passengers in the vehicle, we need to define events that reduce comfort. In this work we have focused on four types of events, namely, fast acceleration, hard braking, hitting potholes on a road, and driving thru curves on high speed.

Each of these events has a different signature on readings from the IMU sensors implemented in the smartphones. In the first step, we had to find threshold values to trigger the detection of the given event. This has been done by test drives with multiple passengers while the driver was driving the same stretch of the road with a different pace. In Fig. 6 the thresholds marked by 5 passengers for acceleration speed that makes them uncomfortable is shown.

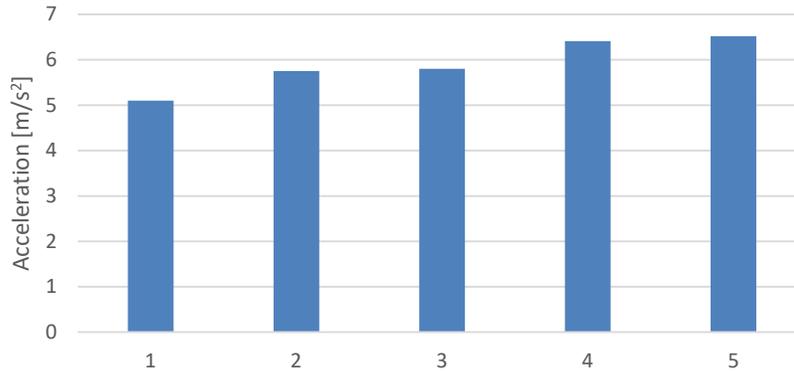

**Fig. 6.** Thresholds for acceleration marked by 5 passengers

From the figure, it is clear that passengers provided different acceleration levels as thresholds for the comfort level, which was expected. However, the settings of the system should consider the lower value of the threshold since, to provide the best results in general. Therefore, the threshold for acceleration was set to 5 m/s².

In the next scenario impact of the turning was examined, input provided by passengers is shown in Fig. 7.

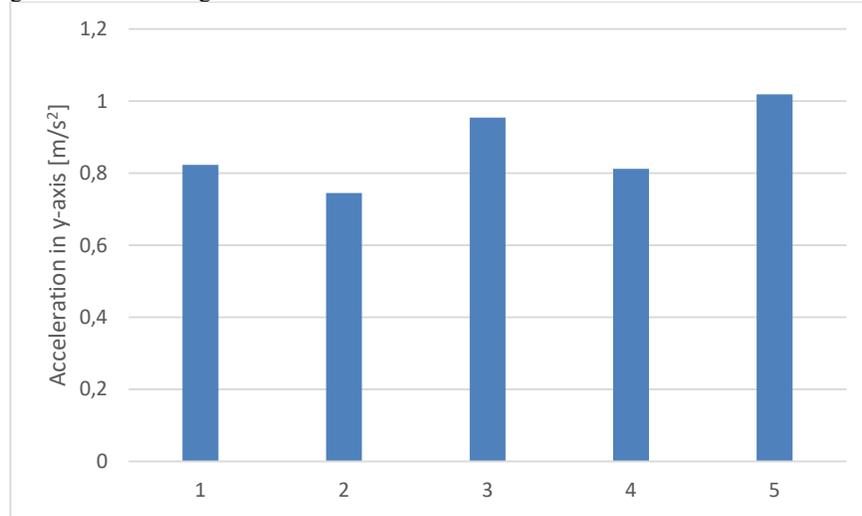

**Fig. 7.** Thresholds for acceleration on y-axis marked by 5 passengers

From the figure, it is obvious that in y-axis significantly lower acceleration is required to lower the comfort level of passengers. In this case, the acceleration threshold was set to 0.75 m/s$^2$.

## 5    Conclusion

In the paper development of a smartphone-based system for automatic evaluation of passenger comfort was described. The system is using data from the IMU implemented in all currently used smartphones. The developed app was tested on multiple passengers to set up thresholds for event detection. However, different passengers reported different threshold values for accelerations on both x and y axes. Therefore, the lowest value was set as a threshold in order to satisfy the preferences of most of the passengers.

In the future we will perform more tests aimed at detection of potholes, that should be visible on changes of acceleration at the vertical axis.

### Acknowledgement

This work has been partially supported by the Slovak VEGA grant agency, Project No. 1/0626/19 "Research of mobile objects localization in IoT environment" and European Union's Horizon 2020 research and innovation programme under the Marie Skłodowska-Curie grant agreement No 734331.